\documentstyle[epsfig]{mn}

\def\tfrac#1#2{{\textstyle\frac{#1}{#2}}}

\title{Filtering techniques for the detection of Sunyaev-Zel'dovich
clusters in multifrequency CMB maps}
\author[Herranz et al.]
       {D. Herranz$^{1,2,}$\footnotemark,
	J.L. Sanz$^{1}$, 
	M.P. Hobson$^{3}$,
	R.B. Barreiro$^{3}$ \footnotemark 
	J.M. Diego$^{4}$,\newauthor
	E. Mart\'\i nez-Gonz\'alez$^{1}$
	and A.N. Lasenby$^{3}$ \\
 $^{1}$Instituto de F\'\i sica de Cantabria,  
             Fac. de Ciencias, Av. de los Castros s/n, 
             39005-Santander, Spain \\
 $^{2}$Departamento de F\'\i sica Moderna, 
             Universidad de Cantabria,       
             39005-Santander, Spain \\
 $^{3}$Astrophysics Group, Cavendish Laboratory, Madingley Road, Cambridge
CB3 0HE, UK \\
 $^{4}$Astrophysics Dept., Keble Road, Oxford OX1 3RH, UK}
\date{}

\pagerange{}
\pubyear{2002}

\begin{document}

\maketitle

\begin{abstract}
The problem of detecting Sunyaev-Zel'dovich (SZ) clusters in
multifrequency CMB observations is investigated using a number
of filtering techniques.  
A multifilter approach is introduced, which optimizes the
detection of SZ clusters on microwave maps. An alternative method is
also investigated, in which maps at different frequencies are combined
in an optimal manner so that existing filtering techniques can be
applied to the single combined map.  The SZ profiles are approximated
by the circularly-symmetric template $\tau (x) = [1 +
(x/r_c)^2]^{-\lambda}$, with $\lambda \simeq \tfrac{1}{2}$ and $x\equiv
|\vec{x}|$, where the core radius $r_c$ and the overall amplitude of
the effect are not fixed a priori, but are determined from the data.  
The background emission is
modelled by a homogeneous and isotropic random field, characterized by
a cross-power spectrum $P_{\nu_1 \nu_2}(q)$ with $q\equiv |\vec{q}|$.
The filtering methods are illustrated by application to simulated
Planck  observations of a $12.8^\circ \times 12.8^\circ$ patch of sky
in 10 frequency channels.  Our simulations suggest that
the Planck instrument should detect $\approx 10000$ SZ clusters in $2/3$ of
the sky. Moreover, we find the catalogue to be complete for fluxes $S
> 170$ mJy at 300 GHz.  
\end{abstract}


\section{INTRODUCTION} \label{introduction_mf}
\footnotetext{e-mail: herranz@ifca.unican.es}
\footnotetext{Currently at Instituto de F\'\i sica de Cantabria,  
             Fac. de Ciencias, Av. de los Castros s/n, 
             39005-Santander, Spain} 
The detection and characterisation of the Sunyaev-Zel'dovich (SZ) 
effect is currently an area of considerable interest 
in millimetre and sub-millimetre astronomy. The SZ effect distorts
primordial cosmic microwave background (CMB) radiation in
the direction of galaxy clusters due to inverse Compton scattering
of CMB photons by the hot intracluster plasma.
In the context of the ESA Planck mission
(and other CMB experiements with high sensitivity and high angular
resolution), one must correct
for this distortion (and for other foreground contaminants)
in order to study the primordial anisotropies in the CMB.
Nevertheless, the SZ effect itself is of considerable cosmological
interest. The effect in individual clusters can be
used to study the intracluster medium. Moreover, since the amplitude
of the effect does not depend on redshift, it can also be used as a
means for detecting clusters that would otherwise be unobservable.
Perhaps most importantly for cosmology, a catalogue of the SZ effects
in a large number of clusters, could be used to place constraints on
cosmological parameters independently of those 
resulting from analysis of primordial
CMB anisotropies. Indeed, a goal of the forthcoming Planck mission
will be to produce a full-sky catalogue of SZ effects 
containing several tens of thousands of galaxy clusters. It is
therefore crucial to have a robust
and efficient method for detecting and extracting the SZ effect from
multifrequency maps of the CMB.

The issue of component separation using multifrequency CMB maps 
has been thoroughly studied in the literature. Proposed methods
include Wiener filtering (WF, Tegmark \& Efstathiou 1996, Bouchet et
al. 1997), the maximum-entropy method (MEM, Hobson et al. 1998, 1999), 
fast independent component analysis (FastICA, Maino et al. 2001), Mexican Hat
Wavelet analysis (MHW, Cay\'on et al. 2000, Vielva el al. 2001a), matched
filter analysis (MF, Tegmark \& Oliveira-Costa 1998) and adaptive filter
techniques (AF, Sanz et al. 2001, Herranz et al. 2001a, Herranz et al. 2001b). 
A non-parametric Bayesian approach to detecting SZ clusters 
has also recently been proposed by Diego et al. (2001b).
A comparison between these methods is difficult because they
have different specific purposes, use different sets of assumptions
and the quality of the separation varies under different
circumstances. For example,
the MHW is designed to detect compact sources with a Gaussian profile (such
as point sources convolved with a Gaussian beam) whereas other methods 
are better suited to deal with diffuse components such as dust and
synchrotron emission. This suggests the 
possibility of combining several of these
methods in order to improve the component separation, for example
MEM $+$ MHW (Vielva et al. 2001b). 

In general, a component separation method that uses all the
available information will be more powerful
than one that does not assume any prior knowledge about the data. 
The MEM, for example, produces excellent results 
when the power spectra and the frequency dependences of the components are well
known. Unfortunately, if these assumptions about the data are
incorrect, errors may
arise that would affect the separation of several (or all) components.
This is particularly dangerous in methods that perform the
separation of all the components
simultaneously (WF, MEM, FastICA). The opposite approach is to
use a robust method that makes as few assumptions as possible about
the data and aims to separate out just a single physical component. 
An example of the latter approach is the non-parametric
SZ detection method given by Diego et al. (2001b),
in which only the well-known frequency dependences of the SZ effect 
and the CMB are assumed. 
In general, however, such methods are less powerful than ones which 
assume a greater degree of (correct) prior knowledge about the data and the
physical component of interest. Clearly, some compromise
between robustness and effectiveness must be made when proposing
a component separation method.

Filtering techniques (such as MHW, MF and AF) are single component 
separation methods that use some of the available information (i.e.
the spatial structure of the component to be detected  
and the power spectrum of the combination of the other `background' 
components), and generally reduce inaccuracies in the separation
that may appear due to error propagation in methods that separate all
components simultaneously. The MHW assumes 
a specific shape (a Gaussian) 
for the component to be separated (`sources')
and, given the power spectrum of the  
background (that can be directly estimated from the data), finds the
optimal scale of the filter. The MF generalises by allowing more
general spatial profiles for the component (usually discrete objects)
to be separated.
Adaptive filters put additional emphasis on
the characteristic scale of the sources in order to reduce 
further the number of spurious detections. Although one usually
assumes spherical symmetry of the sources to be detected, it is not a 
general requirement of filtering method and filters can be easily generalised
to detect asymmetric features.

In this paper, we discuss the generalisation of filtering techniques,
in particular MF and AF, to the case of multiple data maps
corresponding to different frequency channels. Multifrequency
information can be used both to increase the signal of the sources and
to reduce the contribution of the background (`noise').  This
information can be used prior to the filtering of the images (by using
the correlations between the different channels to find an optimal
combination of channels that maximises the signal-to-noise ratio of SZ
clusters) or can be included directly in the construction of the
filters. The structure of this paper is as follows.
In section~\ref{formalism_mf} we describe the formal aspects of the 
 methods that make use of multifrequency information
proposed in this paper. 
In section~\ref{simulations_mf} we describe the simulations we made
to test the multifilters. 
Section~\ref{results_mf} is devoted to summarise and discuss 
our results.

\section{Multifrequency filtering} \label{formalism_mf}

There are two different approaches we can follow to include
the frequency dependence of a signal in the filtering
of multi-channel data.
On the one hand, we can filter each channel separately, but carefully
taking into account the cross-correlation between the different channels
and the frequency dependence of the signal in order to 
obtain an output set of filtered maps that, when added to each other, 
is optimal for the detection. This philosophy gives birth
to the \emph{multifilter method}. On the other hand, 
one can use the information about correlations and frequency
dependence before filtering in order to find the optimal
combination of channels that maximizes the signal-to-noise ratio
of the sources, and then use a filter on the optimally combined
map. This leads to the design of a \emph{combination method}
plus a single filtering. These two methods are discussed in detail
below. We first, however, define our model of the multifrequency
observations.

\subsection{Model of the observations}

Let us consider a set of 2-dimensional images (maps) 
with data values defined by
\begin{equation} 
y_{\nu}(\vec{x}) = f_{\nu}s_{\nu}(x) + n_{\nu}(\vec{x}),\ \ \ 
x\equiv |\vec{x}|,\ \ \ \nu = 1,...,N,
\label{freqmaps}
\end{equation}
where $N$ is the number of maps (or number of frequencies).
The first term on the RHS represents the {\em signal} in each frequency
map due to the thermal SZ effect in clusters, and
the {\em generalised noise} $n_\nu(\vec{x})$ corresponds to the sum of the 
other emission components in the map. 

For a map of any reasonable size, 
we would expect the presence of several SZ clusters. To
illustrate more clearly the construction of multifilters, however,
we assume that the signal is due to a single SZ cluster at the origin
$\vec{x}=0$ of the map (the generalisation to several clusters
distributed across the map is straightforward). In particular,
we assume a spherically-symmetric 
$\beta$-profile for the 
cluster electron number density 
\[
n_e(r)\propto [1 + (x/r_c)^2]^{-\frac{3}{2}\beta},
\]
where $r_c$ is the core radius and we adopt the standard value 
$\beta = 2/3$. One trivially obtains that
the two-dimensional microwave decrement from such a cluster has the
form $A\tau(x)$, where $A$ is the amplitude of the effect and 
$\tau(x)$ is the spatial template
\begin{equation} 
\tau (x) = \frac{1}{[1 + (\frac{x}{r_c})^2]^{1/2}}.
\end{equation}
At each observing frequency this template is convolved with the
corresponding antenna beam, which we assume to be a 2-D Gaussian
of dispersion $\theta_\nu$, to produce the convolved profile
$\tau_{\nu}(x)$. Thus, in (\ref{freqmaps}), 
$s_{\nu}(x) \equiv A\tau_{\nu}(x)$. The quantity 
$f_{\nu}$ in (\ref{freqmaps}) is the frequency 
dependence of the SZ effect, normalised such that
$f_{{\nu}_f} = 1$ at the fiducial
frequency $\nu_f$. Hence $A$ is the
true amplitude of the SZ effect at the frequency $\nu_f$.

The background $n_{\nu}(\vec{x})$ is modelled as a homogeneous 
and isotropic random field with average value 
$\langle n_{\nu}(\vec{x})\rangle = 0$ and cross-power spectrum 
$P_{{\nu}_1 {\nu}_2}(q)$ ($q\equiv |\vec{q}|$) defined by
\[
\langle n_{\nu_1}(\vec{q})n^*_{\nu_2}(\vec{q'})\rangle = 
P_{\nu_1\nu_2}(q) \, \delta_D^2 (\vec{q} - \vec{q'}),
\]
where $n_{\nu}(\vec{q})$ is the Fourier transform of 
$n_{\nu}(\vec{x})$ and $\delta_D^2$ is the 2-D Dirac distribution.

\subsection{The mutlifilter method}

We now consider the first of two approaches to the filtering of
mutlifrequency data, namely the multifilter. In fact, multifilters
themselves can be constructed in several different ways. We consider
here the two main possibilities, which are the {\em scale-adaptive
multifilter} and the {\em matched multifilter}.

\subsubsection{Scale-adaptive multifilter (SAMF)} \label{section_samf}

The idea of an optimal scale-adaptive filter was
recently proposed by Sanz et al. (2001), which we now generalise
to multifrequency data. One begins by
introducing, for each observing frequency, a circularly-symmetric
function $\psi_\nu(x)$. From this function, one can define
a family of filter functions
\begin{equation} \label{SAMF}
\Psi_{\nu} (\vec{x}; R_{\nu}, \vec{b}) = 
\frac{1}{R^2}\psi_{\nu} \left(\frac{|\vec{x} - \vec{b}|}{R_{\nu}} \right),
\end{equation}
where the 3 parameters $R_{\nu}$ and $\vec{b}$ define a scaling and
a translation respectively. For any particular values of these
parameters, we define the filtered map at the $\nu$th observing
frequency by
\begin{equation} 
w_{\nu}(R_{\nu}, \vec{b}) = \int d\vec{x}\,y_{\nu}(\vec{x})
\Psi_{\nu} (\vec{x}; R_{\nu}, \vec{b}).
\label{convol}
\end{equation}
and the `total' filtered map as
\begin{equation} 
w(R_1,...,R_n, \vec{b}) \equiv \sum_{\nu} w_{\nu}(R_{\nu}, \vec{b}).
\label{total}
\end{equation}

The convolution (\ref{convol}) can be written as a product in 
Fourier space, in the form
\begin{equation} 
w_{\nu}(R_{\nu}, \vec{b}) = \int d\vec{q}e^{- i\vec{q}\cdot \vec{b}}
\,y_{\nu}(\vec{q})\psi_{\nu}(R_{\nu}q),
\end{equation}
where $y_{\nu}(\vec{q})$ and $\psi_{\nu}(q)$ are the Fourier transforms 
of $y_{\nu}(\vec{x})$ and $\psi_{\nu}(\vec{x})$, respectively. 
A simple calculation then shows that
the expectation value of the filtered field at the origin 
$\vec{b} = 0$, is given by
\begin{equation} 
\langle w_{\nu}(R_{\nu}, \vec{0})\rangle 
= 2\pi A\,f_{\nu}\int dq\,q\tau_{\nu}(q)
\psi_{\nu}(R_{\nu}q),
\end{equation}
where the ensemble-average is over realisations of the background
emission $n_\nu(\vec{x})$ and the limits in the integrals go from 
$0$ to $\infty$. Similarly, one finds that the variance
of the total filter field (\ref{total}) is given by
\[
\sigma_w^2(R_1,...,R_n) \equiv 
\langle w^2(R_1,...,R_n, \vec{b})\rangle - 
{\langle w(R_1,...,R_n, \vec{b})\rangle}^2 
\]
\begin{equation}
\qquad\qquad =  2\pi \sum_{\nu_1 ,\nu_2}  \int dq\,q\,P_{\nu_1 \nu_2}(q)
\psi_{\nu_1}(R_{\nu_1}q)\psi_{\nu_2}(R_{\nu_2}q).
\label{vardef}
\end{equation}

We choose the filter functions $\psi_\nu(q)$ to optimise
the detection of the cluster at the origin,  taking into
account that the source has a bell shape with a single characteristic scale 
$R_{o \nu}$ in each map. We define the optimal scale-adaptive
multifilter (SAMF) to be that which ensures the following
conditions are satisfied:
\begin{enumerate}
\item{$w(R_{o1},..., R_{on}, \vec{0})$ is an {\em unbiased} estimator 
of the amplitude of the source, so
$\langle w(R_{o1},..., R_{on}, \vec{0})\rangle = A$;}
\item{the variance of $w(R_1,...,R_n, \vec{b})$ has a minimum at the scales 
$R_{o1},...,R_{on}$, i.e. it is an ${\it efficient}$ estimator;}
\item{$w_{\nu}(R_{\nu}, \vec{b})$ has a maximum at $(R_{o\nu}, \vec{0})$.}
\end{enumerate}
The multifilter satisfying these conditions is given by the matrix equation
\begin{equation} 
\tilde{\psi} (q) = P^{-1}(\alpha F + G),
\label{samfdef}
\end{equation}
where we have introduced the column vectors
$\tilde{\psi} (q) = [\psi_{\nu}(R_{\nu}q)]$, $F = [f_{\nu}\tau_{\nu }]$,
and $G = [\mu_{\nu }\beta_{\nu }]$, where
\[
\mu_{\nu }\equiv f_{\nu}\tau_{\nu }\left(2 + 
\frac{d\ln\tau_{\nu}}{d\ln q}\right).
\]
Also, $P^{-1}$ is the inverse of the matrix $P\equiv [P_{\nu_1
\nu_2}(q)]$, and the quantities $\alpha$ and $\beta_\nu$ are given by
the components
\begin{equation} 
\alpha = (A^{-1})_{00},\qquad \beta_{\nu} = (A^{-1})_{\nu 0},
\end{equation}
where $A$ is the $(1+n)\times (1+n)$ matrix with elements
\begin{equation} 
A_{00}\equiv \int d\vec{q}\,F^tP^{-1}F,\quad
A_{0{\nu}}\equiv \int d\vec{q}\,\mu_{\nu }(F^tP^{-1})_{\nu },
\end{equation}
\begin{equation} 
A_{\nu 0}\equiv \int d\vec{q}\,\mu_{\nu}(P^{-1}F)_{\nu},\quad
A_{\nu\nu^\prime}\equiv \int d\vec{q}\,\mu_\nu \mu_{\nu^\prime}
(P^{-1})_{\nu\nu^\prime}.
\end{equation}
The quantity $\tilde{\psi} (q)$ in (\ref{samfdef})
is called {\em scale-adaptive} multifilter, extending the concept 
considered in our previous paper (Sanz et al. 2001) for a single map.
Also, from (\ref{vardef}), we see that the variance can be written as
\begin{eqnarray}
{\sigma}^2_w & = & \int d\vec{q}\,{\tilde{\psi}}^tP\tilde{\psi}
\nonumber \\
& = & (A_{00})^{-1}\left[1 + 2\int d\vec{q}\,F^tP^{-1}G\right] + 
\int d\vec{q}\,G^tP^{-1}G. \nonumber
\end{eqnarray}

An interesting special case is when $P$ is a diagonal matrix, i.e. 
there is no cross-correlation between the backgrounds 
in the different frequency maps, so $P_{\nu \nu^{\prime}} = 
{\delta}_{\nu {\nu }^{\prime}}P_{\nu }(q)$. In this case, 
the multifilter is given by
\begin{equation} 
{\tilde{\psi}}_{\nu} (q) = \frac{f_{\nu}\tau_{\nu}}{P_{\nu }}
\frac{1}{a - \sum_{\nu}f_{\nu}^2 \frac{b^2_{\nu }}{H_{\nu }}} 
\left[1 - \left(2 + \frac{d\ln\tau_{\nu}}{d\ln q}\right)
\frac{b_{\nu }}{H_{\nu }}\right], 
\end{equation}
where
\begin{eqnarray}
a & = & \sum_{\nu}f_{\nu}^2\int d\vec{q}\,\frac{\tau^2_{\nu}}{P_{\nu }},\\
b_{\nu } & = & f_{\nu}^2\int d\vec{q}\,\frac{\tau^2_{\nu}}{P_{\nu }}
\left(2 + \frac{dln\tau_{\nu}}{dlnq}\right), \\
H_{\nu } & = & f_{\nu}^2\int d\vec{q}\,\frac{\tau^2_{\nu}}{P_{\nu }}
{\left(2 + \frac{d\ln\tau_{\nu}}{d\ln q}\right)}^2.
\end{eqnarray}
If, additionally, one assumes that the backgrounds are 
white noise, i. e. the $P_{\nu}$ are
constants and the resolution is the same in all maps, 
producing the convolved profile
$\tau (r_c, \theta )$, then
\begin{equation}
f_{\nu}{\tilde{\psi}}_{\nu} (q)  =  \frac{\tau}{N}
\frac{f^2_{\nu}P^{-1}_{\nu}}{\sum_{\nu }f^2_{\nu}P^{-1}_{\nu}}
\left[1 - \frac{b}{H}\left(2 + \frac{d\ln\tau}{d\ln q}\right)\right], 
\end{equation}
where
\begin{eqnarray}
N & \equiv & \int d\vec{q}\,{\tau}^2 - \frac{b^2}{H}, \\
b & = & \int d\vec{q}\,{\tau}^2\left(2 + \frac{d\ln\tau}{d\ln q}\right), \\
H & = & \int d\vec{q}\,{\tau}^2{\left(2 + \frac{d\ln\tau}{d\ln q}\right)}^2.
\end{eqnarray}

\subsubsection{Matched multifilter (MMF)}

If we do not assume condition (iii) in the definition of the SAMF 
in the previous subsection, one obtains
another type of multifilter after minimization of the variance
condition (ii) subject to the constraint (i).
The multifilter satisfying these less restrictive 
conditions is given by the matrix equation
\begin{equation} 
\tilde{\psi} (q) = \alpha P^{-1}F ,\qquad 
{\alpha}^{-1} = \int d\vec{q}\,F^tP^{-1}F,
\end{equation}
where $F$ is the column vector $F = [f_{\nu}\tau_{\nu }]$ and $P^{-1}$ is the
inverse matrix of the cross-spectrum $P$. This will be called {\em matched}
multifilter extending the concept usually considered for a single map.
In this case, the variance of the filtered field is given by
\begin{equation} 
{\sigma}^2_w = \int d\vec{q}\,{\tilde{\psi}}^tP\tilde{\psi} = \alpha.
\end{equation}

In the special case where there is no cross-correlation
between the backgrounds in the different maps, so $P_{\nu \nu^{\prime}} = 
{\delta}_{\nu {\nu }^{\prime}}P_{\nu }(q)$, the multifilter is given by
\begin{equation} 
{\tilde{\psi}}_{\nu} (q) = \alpha f_{\nu}\tau_{\nu}P^{-1}_{\nu },\qquad
{\alpha}^{-1} 
=  \sum_{\nu}f_{\nu}^2\int d\vec{q}\,\frac{\tau^2_{\nu}}{P_{\nu }}.
\end{equation}
If additionally, one assumes that the backgrounds are white noise, 
i.e. the $P_{\nu }$ are
constants, and the resolution is the same in all maps, producing 
the convolved profile $\tau (r_c, \theta )$, then
\begin{equation} 
f_{\nu}{\tilde{\psi}}_{\nu} (q) = \frac{\tau}{N}
\frac{f^2_{\nu}P^{-1}_{\nu}}{\sum_{\nu }f^2_{\nu}P^{-1}_{\nu}},\qquad
N\equiv \int d\vec{q}\,{\tau}^2.
\end{equation}

A similar result has been independently developed 
by Naselsky et al. (2001).

\subsection{The combination method} 

Our second approach to the filtering of mutlifrequency data
is to perform a two stage process in which one first constructs
some optimal linear combination of the individual frequency maps, and
then applies a single filter to this combined map.

The linear combination of the original frequency maps
is constructed so that it maximally 
`boosts' objects with a given spatial template and
frequency dependence above the background. In general, the
combination map is given by
\begin{equation}
y(\vec{x}) = \sum_{\nu }c_{\nu}y_{\nu}(\vec{x}),
\label{map_combination}
\end{equation}
where $c_\nu$ is some set of weights (derived below). From (\ref{freqmaps}),
we see that this combined map can be written as
\begin{equation}  
y(\vec{x}) = At(x) + \epsilon (\vec{x}),
\label{singlemap} 
\end{equation}
where the `combined' template $t(x)$ and background
$\epsilon(\vec{x})$ are given simply by
\begin{equation}
t(x) = \sum_{\nu }c_{\nu}f_{\nu}\tau_{\nu}(x), \qquad
\epsilon (\vec{x}) = \sum_{\nu }c_{\nu}n_{\nu}(\vec{x}).
\end{equation}
The optimal values of the weights $c_\nu$ are found by maximising
the quotient $Q = At(\vec{0})/{\sigma}_{\epsilon}$, where 
${\sigma}_{\epsilon}$ is the dispersion of the combined noise
field $\epsilon (\vec{x})$. Thus, one is maximising the
`detection level' of the object of interest above the background.

In practice, it is easier (but equivalent) to maximise $Q^2$.
It is straightforward to show that, in the numerator of $Q^2$,
\begin{equation}
[t(\vec{0})]^2 = c^tGc,
\end{equation}
where 
\begin{equation}
c =(c_{\nu}), \quad G = (G_{\nu \nu^{\prime}}), 
G_{\nu \nu^{\prime}} = 
f_{\nu}\tau_{\nu}(0)f_{\nu^{\prime}}\tau_{\nu^{\prime}}(0).
\end{equation}
Similarly, in the denominator of $Q^2$,
\begin{equation}
{\sigma}_{\epsilon}^2= c^tMc, 
\end{equation}
where
\begin{equation}
c =(c_{\nu}), \ \ \ M = (M_{\nu \nu^{\prime}}), \quad
M_{\nu \nu^{\prime}} = 
\langle n_{\nu}(\vec{x})n_{\nu^{\prime}}(\vec{x})\rangle.
\end{equation}
Thus one must maximise, with repect to the weights $c_\nu$, the
quantity
\begin{equation}
Q^2 = A^2\frac{c^tGc}{c^tMc}.
\end{equation}
The solution is easily found to be the eigenvector $c$ 
associated to the largest 
eigenvalue $\lambda$ of the generalized eigenvalue problem
\begin{equation}  \label{eigenvalue} 
(G - \lambda M)c = 0.
\end{equation}

Once we have constructed the combined template $t(x)$, given in 
(\ref{map_combination}), with the optimal
weights $c$ given by (\ref{eigenvalue}), we can apply 
a either a scale-adaptive filter or matched filter to the 
{\em single} combined image given by (\ref{singlemap}).
The appropriate form of the scale-adaptive filter for a single image
is given by Sanz et al. (2001). The appropriate form of the matched filter
for a single image is given by
\begin{equation}
\tilde{\psi} (q) = \frac{1}{2\pi a}\frac{t(q)}{P(q)}, 
\end{equation}
where
\begin{equation}
a = \int dq\,q\frac{t^2(q)}{P(q)}, \quad
P(q) = c^tNc, \quad
N = [P_{\nu \nu^{\prime}}(q)],
\end{equation}
where $t(q)$ and $P(q)$ are the Fourier transform of $t(x)$ and the 
power spectrum of the noise $\epsilon$, respectively.

\subsection{Comparison of methods}

We have introduced two different methods for filtering
mutlifrequency data, namely multifilters and the combination
method. Moreover, each method can be implemented 
for two different kinds of filter (scale-adaptive filters and matched
filters). This makes four
different possible ways to filter the data, and we would expect
each method to have advantages and shortcomings with respect to the
others.
 
From the methodological point of view, the main difference
between the combination method plus a single filter and the multifilter method
is that the first one uses the multi-frequency information
to give the {\it optimal starting point} for the filter, while 
multifilters produce the {\it optimal ending point} after
filtering. In that sense, multifilters are more powerful 
than a single filter applied to an optimally combined map. 
The cost of this higher efficiency, however, is an increase in the 
complexity of the filters and therefore of the computational
time required to perform the data analysis. For example,
in the application to simulated Planck observations 
in 10 frequency channels described below, the multifilter method
requires $\sim 10$ times more CPU time than the combination
method (because the last one filters only once).

Regarding the type of filter, matched filters give higher
gains while adaptive filters reduce the number of errors
in the detections (spurious sources). This is due to the
fact that the optimal scale-adaptive filter 
includes a constraint (numbered (iii) above) about the scale of the sources
that is not present in matched filter design. This
constraint characterises with more precision the sources
but also restricts the minimisation of the variance 
in the filtered maps (that is, the final gain).

\section{SIMULATIONS} \label{simulations_mf}

In order to test the multifrequency filtering methods outlined above,
we now apply them to simulated Planck observations.
These simulations mimic the main features of the future Planck
mission, and are realistic in the sense that
they include the latest available information about different
physical components of emission 
(CMB, Galactic components, SZ effect and extra-galactic point
sources) and that they reproduce the technical specifications 
of the different Planck channels (pixel sizes, antenna beams and noise
levels). Table~\ref{tech_mf} shows the assumed observational
characteristics of the simulated maps.
The simulations were performed in patches of the sky of 
$12.8^{\circ} \times 12.8^{\circ}$. However, the method can be easily 
extended to the full sphere.

\begin{table}
\begin{center}
 \begin{minipage}{84mm}	
 \caption{Technical characteristics of the 10 simulated Planck channels.
Column two lists the FWHM assuming a Gaussian beam. Column three shows
the pixel size in arcmin. Column four lists the fractional bandwidths of
each channel. In column five the instrumental noise is $\Delta T$ ($\mu K$)
per resolution element for 12 months of observation is given.}
 \label{tech_mf}
 \begin{tabular}{c c c c c}
\hline
Frequency & FWHM & Pixel size & Fractional  & $\sigma_{noise}$ \\
(GHz)     & (arcmin) & (arcmin) & bandwidth & ($\mu K$) \\
          &      &            &  ($\Delta \nu / \nu$)  &                 \\

\hline

30	& 33.0 	& 6.0 	& 0.20	& 4.4 	\\
44	& 23.0  & 6.0   & 0.20  & 6.5   \\
70	& 14.0  & 3.0   & 0.20  & 9.8   \\
100 (LFI) & 10.0 & 3.0  & 0.20  & 11.7  \\
100 (HFI) & 10.7 & 3.0  & 0.25  & 4.6   \\
143	& 8.0   & 1.5	& 0.25  & 5.5  	\\
217	& 5.5	& 1.5	& 0.25	& 11.7	\\
353	& 5.0	& 1.5	& 0.25	& 39.3	\\
545	& 5.0	& 1.5	& 0.25	& 400.7 \\
857	& 5.0	& 1.5	& 0.25	& 18182 \\
\hline
 \end{tabular}
 \end{minipage}
 \end{center}
\end{table}

The CMB simulation was generated using the
$C_l$'s provided by the CMBFAST code (Seljak \& Zaldarriaga, 1996)
for a spatially-flat $\Lambda$CDM Universe 
with $\Omega_m=0.3$ and $\Omega_{\Lambda}=0.7$
(Gaussian realization).
\begin{figure*}
\includegraphics[angle=0, width=17cm]{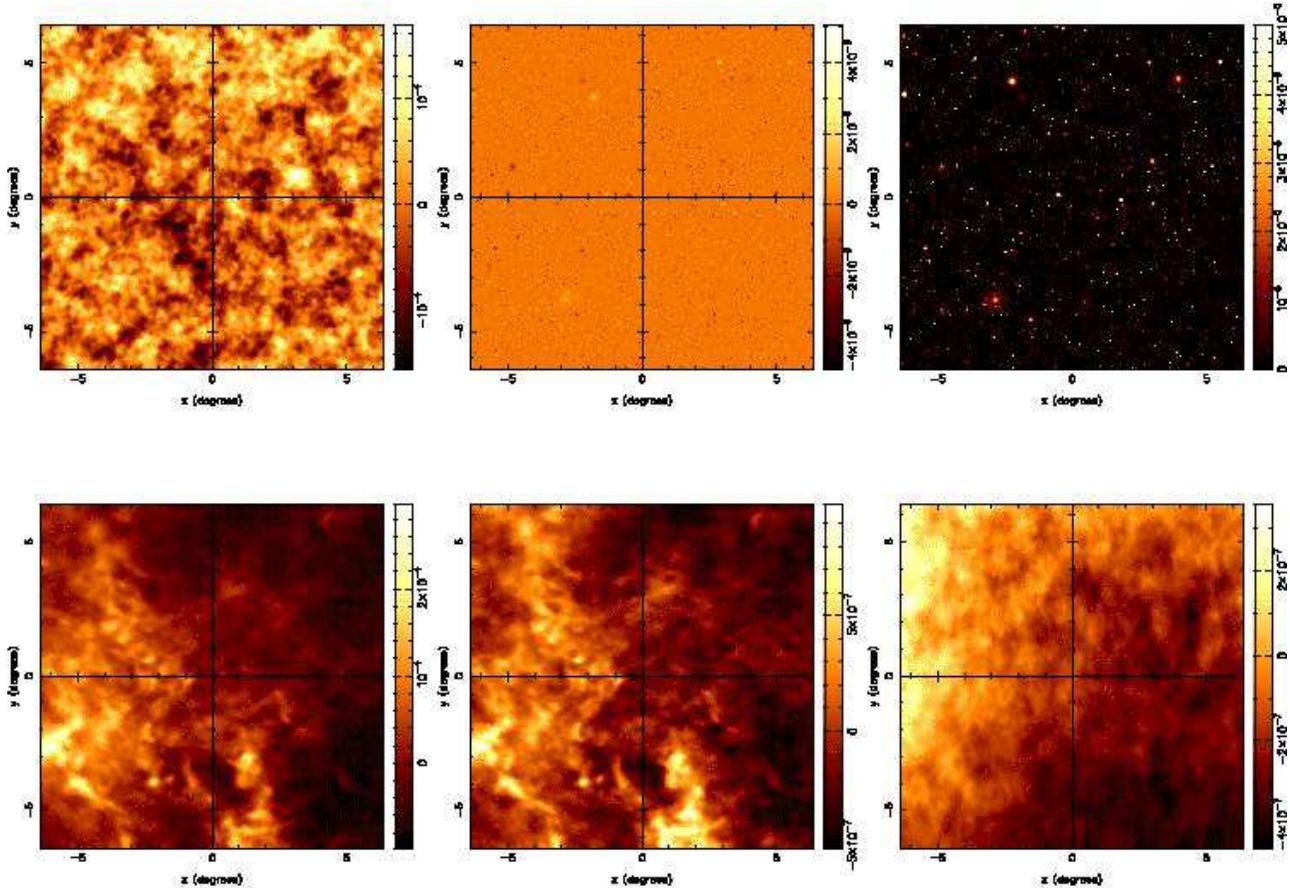}
\caption{Components present in the simulation at 300 GHz. 
From left to right and from top to bottom
the components are: CMB, kinetic SZ, thermal SZ, Galactic dust, free-free
and synchrotron. 
The spinning dust component has the same spatial template as
Galactic thermal dust emission. The units of the maps are $\Delta T/T$.}
\label{inputs}
\end{figure*}
The Galactic emission is assumed to originate from
four different components: thermal dust, spinning dust, free-free and 
synchrotron. Thermal dust was simulated using the template given
by Finkbeiner et al. (1999). 
This model assumes that dust emission is due to two grey-bodies:
a hot one with a dust temperature $T_D^{hot} \simeq 16.2 K$ and an 
emissivity $\alpha^{hot} \simeq 2.70$, and a cold one
with $T_D^{cold} \simeq 9.4 K$ and $\alpha^{cold} \simeq 1.67$. These
quantities are mean values.

For the free-free template we used one correlated with the dust
emission in the manner proposed by Bouchet et al. (1996). 
The frequency dependence 
of the free-free emission is assumed to vary as 
$I_{\nu} \propto \nu^{-0.16}$, and 
is normalised to give an rms temperature fluctuation of 
$6.2\mu K$ at 53 GHZ. 

The synchrotron spatial template has been produced using 
the all sky map of Fosalba \&
Girardino \footnote{ftp://astro.estec.esa.nl/pub/synchrotron}, 
which is an extrapolation of the
408 MHz radio map of Haslam et al. (1982), from the original 1 deg resolution 
to a resolution of 5 arcmin. 
We further extrapolated
the small-scale structure to 1.5 arcmin
following a power-law power spectrum with an exponent of $-3$. 
The frequency 
dependence is assumed to be $I_{\nu} \propto \nu^{-0.9}$ and 
is normalized to the Haslam 408 MHz map. We include in our simulations
information on the changes of the spectral index
as a function of electron density in the Galaxy.
This template has been made by combining the Haslam map with
the Reich \& Reich (1986) map at 1420 MHz and the Jonas et al. (1998) map at
2326 MHz, and can be found in the FTP site refered to above.

We have also taken into account the possible Galactic emission 
due to spinning grains of dust, proposed by Draine \& Lazarian (1998).
This component could be important at the lowest frequencies of
Planck (30 and 44 GHz) in the outskirts of the
Galactic plane.

\begin{figure*}
\includegraphics[angle=0, width=17cm]{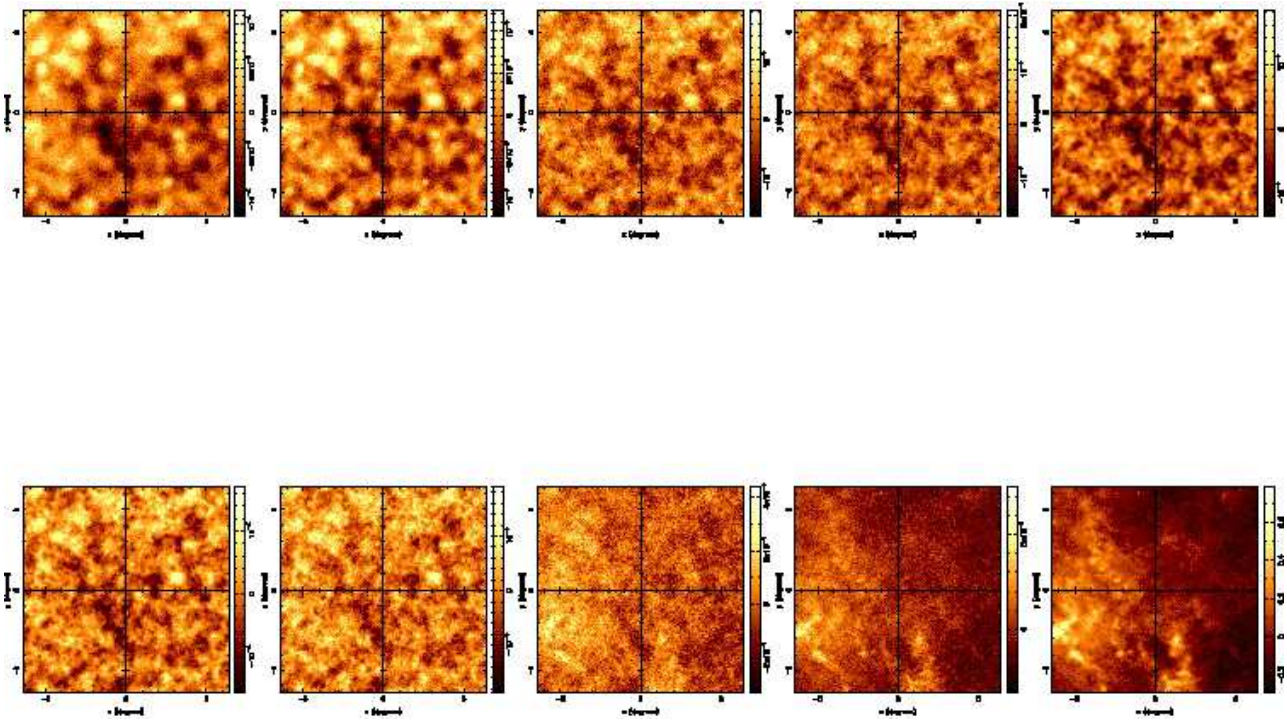}
\caption{Simulated Planck channels. Each map corresponds to the same 
$12.8^{\circ} \times 12.8^{\circ}$ area of the sky at the 
frequency of the channel. The represented channels are, from 
left to right and from top to bottom, 
30, 
30,
44,
70,
100 (LFI),
100 (HFI),
143,
217,
353,
545 and
857 GHz. The units of the maps are $\Delta T/T$.} 
\label{data_Jysr}
\end{figure*}

The extragalactic point source simulations follow the same cosmological model 
as that assumed for the CMB simulation and corresponds to the model of 
Toffolati et al. (1998).
The thermal SZ effect was made for the same cosmological model. 
The cluster population was modelled
following the Press-Schechter formalism (Press \& Schechter 1974) 
with a Poissonian
distribution in $\theta$ and $\phi$. The simulated cluster population fits well
all the available X-ray and optical cluster data sets (see Diego et al. 
2001a for a discussion). The different components used for the simulation
are shown at 300 GHz in figure~\ref{inputs}. 
Figure~\ref{data_Jysr} shows the simulated channels taking into account
all the components, as well as the antenna beam effect and the instrumental
noise.

The simulation described above is the same used by Diego et al. (2001b).
We have chosen this particular simulation in order to compare
results with that work under the same conditions.

\section{RESULTS AND DISCUSSION} \label{results_mf}

\subsection{Testing the methods} \label{test_mf}

Before applying the methods presented in section~\ref{formalism_mf} to
the simulations described in section~\ref{simulations_mf}, let us 
illustrate how they work in a simplified case. 
For the sake of simplicity, the simulations
include the same foregrounds and technical features as
the simulations described in section~\ref{simulations_mf},
except for the fact that the spinning dust component was not included.
Spinning dust is the weakest component and its contribution is 
almost negligible in most Planck channels.
The other main difference with the realistic simulations is
the thermal SZ effect itself. Instead of the realistic SZ clusters,
we simulated 200 clusters of the  same size ($r_c=0.5$ pixels), and
amplitudes uniformly distributed between 0 and the maximum
amplitude of the clusters belonging to the realistic simulation.
The simulated clusters have the frequency dependence of the thermal SZ
effect and the radial profile
\begin{equation}\label{modified_multiquadric}  
\tau(x)=N\left(\frac{1}{\sqrt{r^2_c+x^2}}-\frac{1}{\sqrt{r^2_v+x^2}}\right)
\end{equation}
where $N=r_v r_c/(r_v+r_c)$. Here $r_c$ denotes the core radius and
$r_v$ is a `cut scale' that can 
be interpreted as the virial radius of the cluster.  
The profile given by (\ref{modified_multiquadric}) is a modified 
multiquadric profile that behaves as a $\beta$ model for $x<<r_v$ and decays 
quickly for $x>>r_v$. In order to simplify the simulations still further, 
the kinematic SZ effect is not included. In the real case, 
the contribution of the kinematic SZ effect is expected to be $\sim 30$ 
times lower than the contribution of the thermal SZ effect, thus
justifying our approximation.

\begin{figure*}
\epsfxsize=170mm
\epsffile{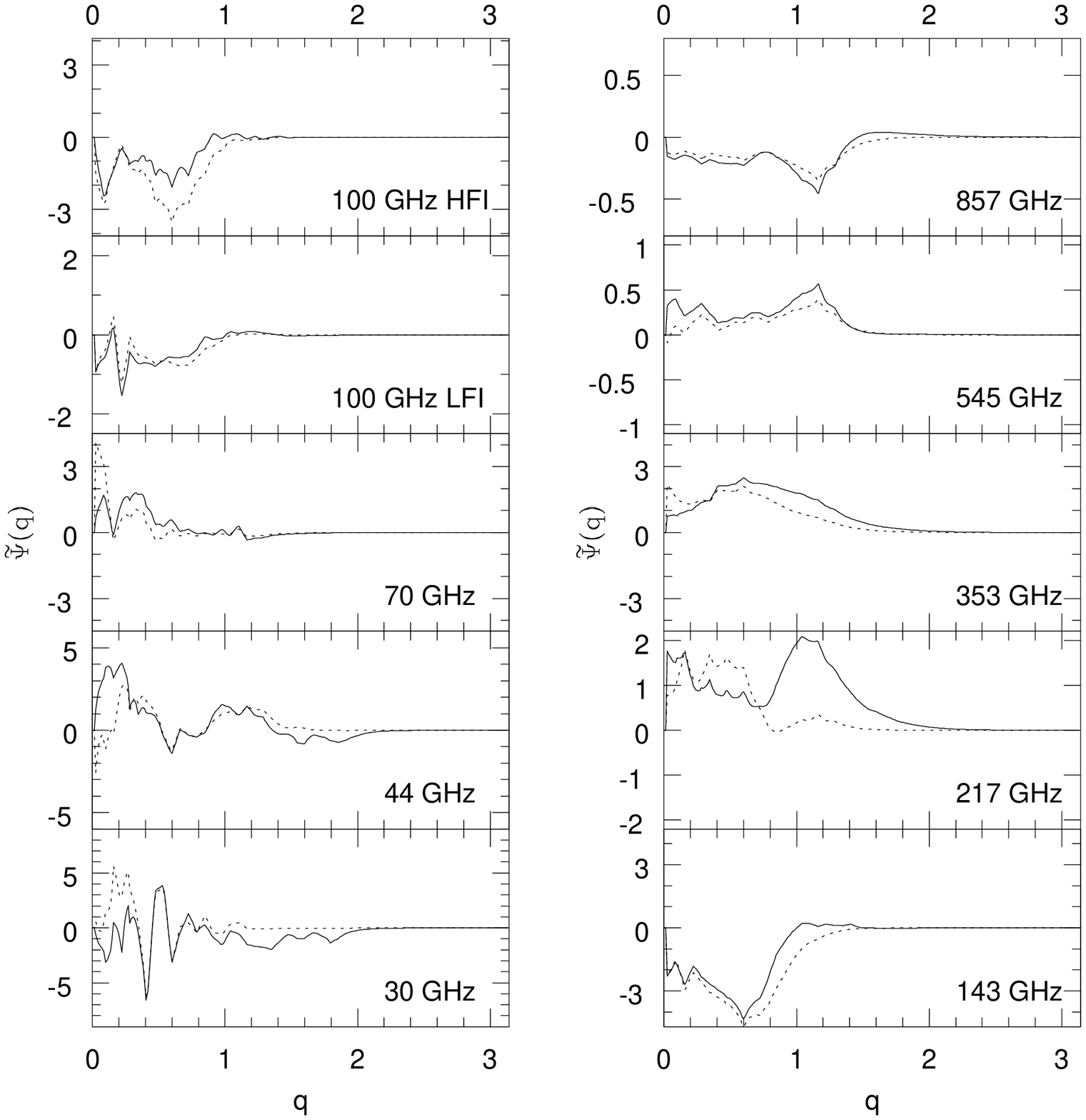}
\caption{Filters used to test the multifilter method. Adaptive filter ({\it solid line})
and matched filter ({\it dotted line}) are represented in Fourier space 
for each frequency channel.}
\label{filters_10freqs}
\end{figure*}

\begin{figure}
\epsfxsize=84mm
\epsffile{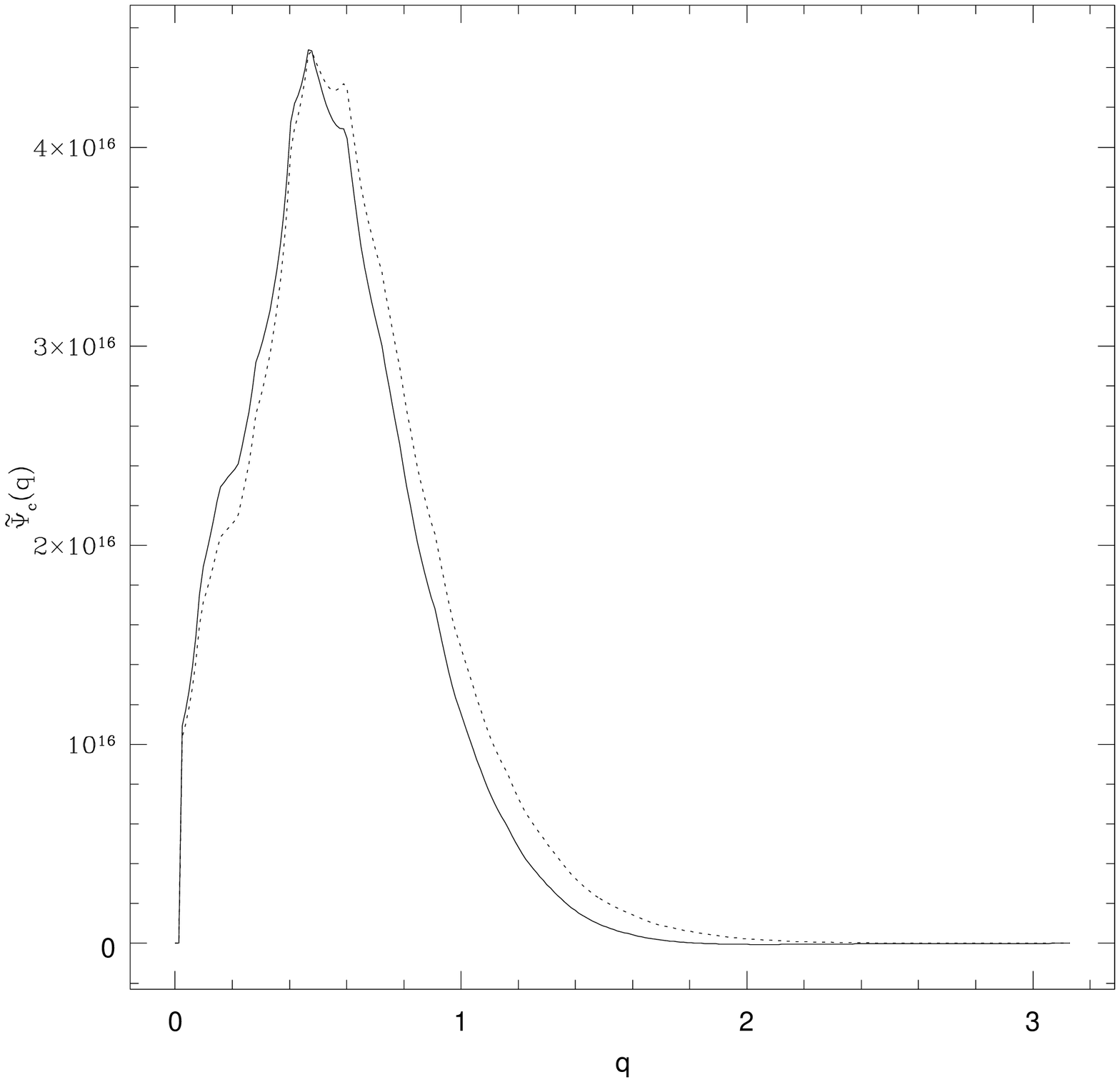}
\caption{Filters used to test the combination method. 
The adaptive filter ({\it solid line})
and the matched filter ({\it dotted line}) are represented in Fourier space.  }
\label{filters_combination}
\end{figure}

The simulated maps were filtered using the four possible combinations
of filters and methods presented in this work. The combination
method took about 1 minute of processing in a 700 MHz PC. The
multifilter method took about 10 minutes of processing on the
same computer. Processing times were slightly higher for the
case of adaptive filters than in the case of matched filters.
To compute the filters and the combinations of the maps the
low resolution channels were re-binned to the resolution
of the highest resolution channel (1.5 arcmin). 
Figure~\ref{filters_10freqs} shows the filters (in Fourier space) 
employed in the
multifilter method. The adaptive filters are represented by solid lines,
whereas matched filters are represented by dotted lines.
Filters, specially those used in low frequency channels, are quite 
complicated and show several peaks at different wave numbers $q$ that
correspond to the different scales the 
filters are trying to identify on the
images.
Conversely, the filters used after the combination method 
are rather simple and are shown in figure~\ref{filters_combination}. 
The differences between matched and adaptive filters in the combination method
are small. This indicates that there will be few differences in the results
of both filters. In the case of the multifilters the differences are 
more pronounced, in particular for the 217 GHz channel.

The results from the test can be summarised as follows.
The number of detections is higher in the multifilter case. This is not
surprising since by definition the multifilter method is more powerful
than the combination method.
In particular, if one raises the detection threshold, 
the difference between the two methods
increases. As an example, table~\ref{results_testmf} shows
the results of one of the tests. After
filtering, the sources were detected by looking for peaks above a $4\sigma$ 
threshold and then compared with a 
catalogue of the original simulated clusters.
By comparison, in the same simulations, if the detection
threshold is raised to $5\sigma$ (this case is not shown 
in table~\ref{results_testmf}) the matched multifilters produce 90
detections whereas the matched filter in the combination method gives only 78.

A detected peak is considered a spurious detection if the distance
to the closest 
object in the original catalogue is greater than 1.5 pixels. The number of
spurious detections is strikingly low in all cases. Nevertheless, 
the adaptive filter seems
to work better than the matched filter (for example, 
in table~\ref{results_testmf} the adaptive filter gives 
0 spurious detections instead of
1 with the combination method). 
Of all the components present in the simulations, 
point sources are the most likely to produce spurious detections due to the
similar scale of sources and SZ clusters. However, the frequency dependence
of the SZ effect greatly reduces the probability of contamination. 

The position of the sources was determined with errors below the pixel
size (1.5 arcmin). However, the error in the determination 
of the amplitude is quite large ($\sim 35 \%$).
Table~\ref{results_testmf} shows that this error is due to a systematic bias.
To explain this bias let us consider the case of matched filter.
The filter normalization is given by integral $a=\int^{\infty}_{0}dqq\tau^2/P$,
being $\tau$ the source profile in Fourier space and $P$ the power spectrum of
 the background.
However, when we analyse a patch of the sky we do not have
information about the power spectrum at all the wave numbers $q$. An 
image divided into pixels is limited by a minimum value $q_{min}$ and
a maximum $q_{max}$. Therefore, the normalization we calculate is
$a'=\int^{q_{max}}_{q_{min}}dqq\tau^2/P$. For the case of 
images with the same size and pixel scale as our simulations, and 
a multiquadric
profile $\tau$ with $r_c=0.5$ pixels, 
the normalisations can differ by $10\%$ for 
a background spectral index $\gamma=1$ to $60\%$ for 
a background spectral index $\gamma=3$. The case for 
adaptive filters is more complicated but qualitatively similar. 
This bias is independent of the source flux and can be calibrated 
using simulations.

\begin{table*}
\begin{center}
 \begin{minipage}{120mm}	
 \caption{Results of the test for 10 frequency 
channels and clusters of equal size 
($r_c=0.5$ pixels). The detection was performed 
over the $4\sigma$ threshold in all
cases. The filtering method is listed in the first 
column. Second column shows the number
of true detections found over the $4\sigma$ detection 
threshold. The third column indicates the 
number of spurious sources detected in each case. The 
fourth column lists the mean error in the
determination of the position of the detected sources. 
Column five lists the mean relative
bias in the determination of the amplitude, 
defined as $\bar{b}_A=100  <(A-A_0)/A_0>$. Column six 
shows the mean relative error in the determination of the amplitude,
defined as  $\bar{e}_A=100 <\mid A-A_0 \mid /A_0>$.
}
 \label{results_testmf}
 \begin{tabular}{c c c c c c}
\hline
METHOD                 & number of  & number of & mean offset & $\bar{b}_A (\%)$ & $\bar{e}_A$ (\%) \\
                       & detections & spurious  & (pixels)    &                  &         \\
\hline
combination/matched    &   110      &    1      &  0.41       &   33.3           &    33.3     \\
combination/adaptive   &   113      &    0      &  0.38       &   32.5           &    32.5     \\
multifilter/matched   &   116      &    1      &  0.42       &   29.8           &    29.9     \\
multifilter/adaptive  &   109      &    1      &  0.57       &   34.4           &    34.5     \\
\hline
 \end{tabular}
 \end{minipage}
 \end{center}
\end{table*}

\begin{figure}
\epsfxsize=84mm
\epsffile{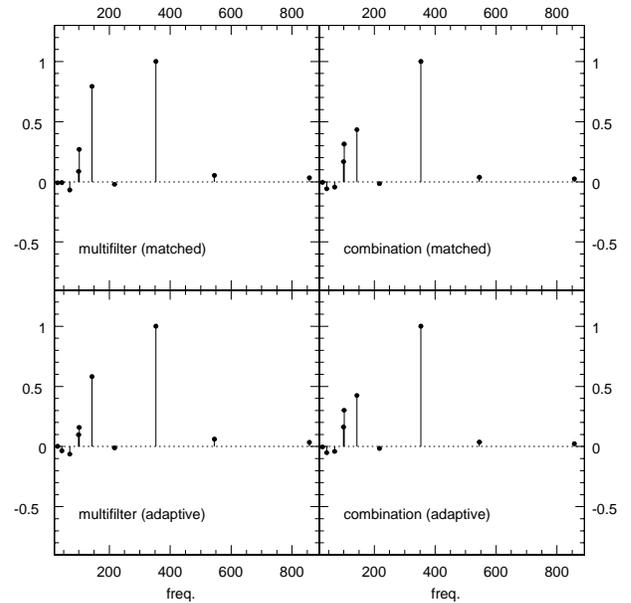}
\caption{Contributions of the different channels to the source amplitude 
estimation. The contributions are normalised so that the channel with 
maximum contribution has contribution 1. Panels corresponding to the combination (matched)
and combination (adaptive) are the same, since the weights are assigned before filtering.
However, they  are  both shown for comparison purposes.}
\label{contributions_4}
\end{figure}

Figure~\ref{contributions_4} shows the contribution of each channel to
the source amplitude estimation for each method. The channels with
larger contribution are the 143 GHz and the 353 GHz. This is not
surprising since they are the channels with more SZ contribution. 
The 100 GHz HFI channel has more contribution than the 100 GHz LFI
because of its better signal-to-noise ratio. 
The combination method puts more emphasis in the 100 GHz 
and less in the 143 GHz channels
than the multifilter method. 
Contributions
from the 30 GHz, 217 GHz and 857 GHz channels are negligible. 
However, that does not mean that these channels do not contribute
to the filter construction. 
For example, if we repeat the analysis with only the 
5 channels with more contribution (70, 100 LFI, 100 HFI, 143 and 353 GHz), 
the number of detections reduces by $\sim 10 \%$ and 
the number of spurious
detections increases to 6 (at $4\sigma$ detection threshold) 
for the combination method case (for the two filters) 
and to 9 (adaptive filter) and
5 (matched filter) for the multifilter method.
Even the 217 GHz channel is important; repeating the analysis with all
channels except for the 217 GHz increases the error 
in the determination of the
amplitudes and also increases the number of spurious detections at low
$\sigma$ thresholds.

We can summarise the conclusions of the test as follows:
\begin{itemize}
\item Multifilter method is more powerful in the detection and estimation of
cluster parameters.
\item Combination method is faster than the multifilter method.
\item Multifrequency information reduces the number of spurious detections. 
Therefore, it is not critical to use an adaptive filter to that end. The 
matched filter allows one to detect more sources.
\item Some channels contribute more than others to the analysis, but all of
them carry valuable information; the analysis should include all the 
available data.
\end{itemize}

\subsection{Results for realistic simulations} \label{results_test_mf}

Taking into account the insights provided by the test presented
in the last sub-section, we are now prepared to 
confront the analysis of realistic
simulations.
These realistic simulations have been described in
section~\ref{simulations_mf}.
The main difference with respect to the previously
performed test is that in the realistic case clusters 
have different sizes that are not known a priori.
Herranz et al. (2001b) proposed a method to 
accommodate with this problem: 
\begin{itemize}
\item Choose a trial core radius $r_c$ and construct the corresponding 
filters. 
\item Convolve the data with the filters, \emph{varying the scales} 
 $R_{\nu}$ \emph{of 
eq.} (\ref{SAMF}). This is performed by 
substituting $\tilde{\psi}(q)$ by $\tilde{\psi}(qx)$, where 
$x=R_{\nu}^{\prime}/R_{\nu}$ is simply a dilation factor.
\item In the case of scale-adaptive filters, condition (iii) 
(see section~\ref{section_samf}) implies that, if the assumed value of
$r_c$ corresponds with the true core radius of the cluster, 
the coefficients will be maximum when $x=1$. If this is not
true, the trial $r_c$ is discarded.
\item In the case of matched filters, condition (iii) no longer
holds.  However, a similar criterion can be used: the best performance of
the filter will occur when the scale of the filter and the scale
of the cluster is the same. Therefore, if the  
signal-to-noise ratio of the coefficients is not maximum for $x=1$, 
the assumed radius $r_c$ is not the optimal choice. If this condition is 
verified for more than one value of $r_c$, the one with the highest
signal-to-noise ratio is chosen.
\item We repeat the process with as many different values of $r_c$
as desired.
\end{itemize}
In Herranz et al. (2001b), the above method has been 
successfully
applied to single frequency maps containing simulated multiquadric
profiles and different kind of backgrounds. 

The clusters in the realistic simulations have a
profile
\begin{equation} \label{realistic_profile}
\tau(r)=\frac{2}{\sqrt{1+(r/r_c)^2}} \tan^{-1} 
\left[\sqrt{ \frac{p^2-(r/r_c)^2}{1+(r/r_c)^2}} \right],
\end{equation}
where $p=r_v/r_c$ is the ratio between the virial radius
and the core radius of the cluster. This profile is realistic
from the point of view of the physics of the Sunyaev-Zel'dovich 
effect. The profiles (\ref{modified_multiquadric}) and
(\ref{realistic_profile}) are almost identical for $r< r_v$. In
order to calculate the filters
in Fourier space,  
it is easier to work with the modified multiquadric 
(\ref{modified_multiquadric}). Therefore, as a first approximation
we will assume that the clusters can be described by the
profile (\ref{modified_multiquadric}). As we will see, this is
a good approximation.

Following the results of the test in the last subsection,
we choose a matched multifilter to perform the analysis.
After applying our method to the simulations we detect
the clusters by looking for sets of connected pixels above
a certain threshold. The maximum of these sets give the position
and the amplitude of the sources.
At the $3\sigma$ level (regions with 5 or more connected pixels)
we are able to detect 63 cluster candidates, of which
62 correspond to real clusters. 
The spurious detection appears in one of the borders of
the image and therefore can be considered as an edge effect.
The mean error in the determination of the position
of the clusters was $\sim 1$ pixel.

\begin{figure*}
\includegraphics[angle=0, width=17cm]{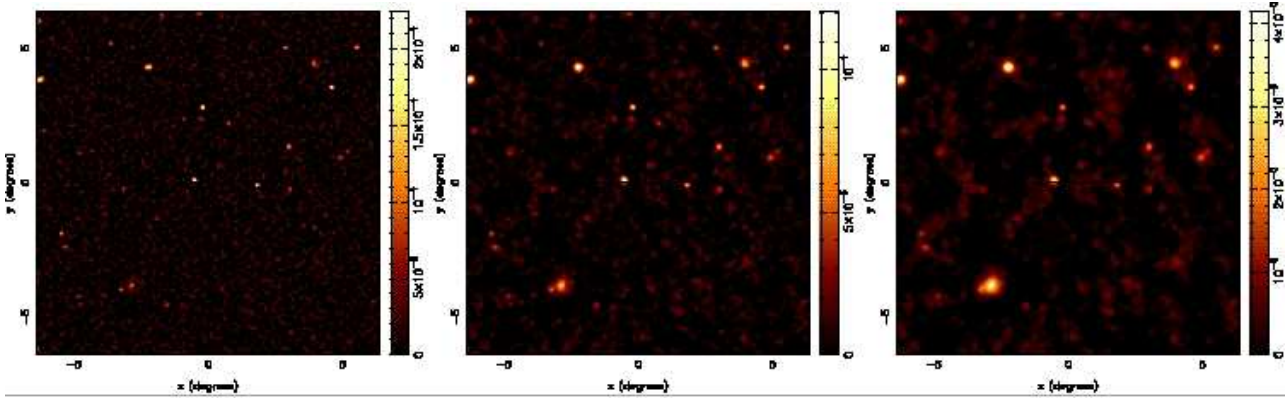}
\caption{Multifiltered data at several scales. Left panel shows 
the output after filtering with a matched multifilter that considers 
that clusters are point-like. Center panel shows the output after filtering
with a matched multifilter for a modified multiquadric profile with $r_c=1.0$
pixel. Right panel shows the output after filtering
with a matched multifilter for a modified multiquadric profile with $r_c=2.5$
pixel. Note how different structures are enhanced at different scales.
The units of the maps are $\Delta T/T$.} 
\label{results_ms}
\end{figure*}

Using the multi-scale analysis we are able to determine the
core radii of the clusters with a mean absolute error 
of 0.30 pixels. The mean bias in the determination of the
core radii was  $-0.15$ pixels. 
Since 
pixelisation effects are expected to corrupt structures
whose typical scale is much smaller than the pixel
size, all clusters with correspondingly small core radii 
will not follow the multi-quadric profile.
Most of the clusters have very
small core radii and therefore can be considered as point-like
sources. 
To detect these clusters, a Gaussian profile (corresponding
to the beam at each channel) was assumed instead of the 
multiquadric given by (\ref{modified_multiquadric}).
The separation between clusters that are considered as point
sources and extended sources was set in $r_c=0.4$ pixel, since 
if $r_c$ is below this limit the FWHM of the
multiquadric profile is of the order of a pixel. All
the clusters that were detected as point sources were considered
to have the (arbitrary) value $r_c=0.1$ pixel.
The bias in the determination of the core radius mentioned above
is due to the asymmetry introduced when assigning
$r_c=0.1$  pixel to all clusters whose core radius is less than 0.4 pixel.
Figure~\ref{results_ms}
shows the resulting maps after filtering with different scales (point source
in the left, $r_c=1.0$ pixel in the middle and $r_c=2.5$ pixel in the right
side). Small clusters stand out mainly in the left panel, while
large clusters are emphasized in the right panel.

\begin{figure}
\epsfxsize=84mm
\epsffile{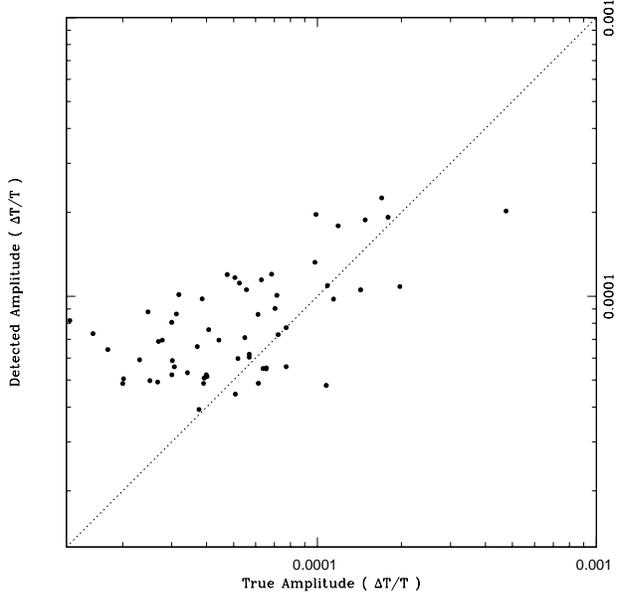}
\caption{Amplitudes of the 60 brightest detected clusters {\it vs.}
true amplitudes. The dotted line corresponds to the 
perfect situation True Amplitude = Detected Amplitude.}
\label{amplitudes}
\end{figure}

Figure~\ref{amplitudes} shows the detected amplitudes of the 
60 brightest detected clusters {\it vs.} 
the true amplitudes. The brightest sources in the graph were recovered
with relative errors of $\sim 30 \%$. As might be expected, 
the error increases as the true amplitude 
decreases. For the weakest clusters, the detected amplitude
tends to a horizontal line. This bias is due to 
the incorrect determination of the normalization 
described in section~\ref{test_mf}, as well as
two additional factors: first,
only the faint clusters that are overamplified are able to
reach the detection threshold; second, the detection is
performed by looking for peaks in the data and, hence, 
those clusters which are by chance enhanced by residual fluctuations
of the background (noise) are more likely to be detected.
The bias due to the normalization should be the same for all
sources with a given scale and therefore could be calibrated 
from simulations, whereas the second kind of bias
affects mainly the weak clusters (that dominate the number counts 
in this case) and can not be easily calibrated.
However, the amplitude estimation could be improved by refining 
the detection method.

The catalogue of detected sources was complete above 0.17 Jy (at 300 GHz).
Below this limit the completeness gradually decreases. There are a few
very faint clusters that are detected with this method (fluxes of
few tens of mJy), but the practical detection limit is $40-50$ mJy
(at 300 GHz). 
The 62 detections in our small sky patch of $(12.8^{\circ})^2$
corresponds to $\approx 10000$ detections in $2/3$ of the sky
(the area not dominated by the Galaxy).

The performance of the other filtering techniques
proposed in this work was compared to the performance
of the matched multifilter for the realistic Planck simulations.
For the comparison, only the point-like
clusters (that dominate the number counts) were considered.
The scale-adaptive multifilter, 
as seen in section~\ref{test_mf},
performs worse than the matched multifilter because their
main advantage, that is the removal of spurious sources,
has already been accomplished by the multi-frequency 
information included in the filters. Since the
clusters one is trying to detect are very faint,
it is the overall gain which is most important.
At $3\sigma$, the scale-adaptive multifilter detects
40 per cent fewer clusters than the matched multifilter.
Both matched and scale-adaptive filtering of an optimally combined map 
are slightly less sensitive than multifilters, 
producing, in both cases, around 20 per cent fewer detections at $3\sigma$  
than with the matched multifilter. In other words,
in the case of single filters, matched and scale-adaptive 
perform very similarly (as was expected 
from figure~\ref{filters_combination}, where 
both filters look alike).

The results
obtained with the matched multifilter 
are comparable in number of detections and reliability 
at the $3\sigma$-threshold 
to
those obtained by Diego et al. (2001b). 
To perform their method it is necessary first to clean
the maps carefully. This cleaning includes
point source removal with a MHW technique 
(Vielva el al. 2001a), dust substraction using the 857 GHz map
and CMB substraction in Fourier space up to a certain
wavenumber limit using 
the 217 GHz map. 
In order to compare results, 
we performed the same cleaning method
and then applied the matched multifilter. At $3\sigma$ 
detection threshold, we obtained a
$7$ per cent more detections than before. At $2\sigma$, we also
detected $7$ per cent more clusters and the number of 
spurious detections decreased by about $12$ per cent. 
This improvement is small given all the cleaning work that is necessary 
to obtain it. 
Moreover, it indicates that the multifilter scheme
is powerful and robust enough to deal successfully 
with contaminants such as point sources, dust and the CMB signal.

\section{Conclusions}

The filtering techniques described in this work
have proven to be useful tools for the detection
of discrete objects in multifrequency
data, which are known to have a prescribed
frequency dependence. The only other assumption
about the objects is their spatial template; any other
quantities of interest are directly calculated
from the data. 
We have assumed in this work that the 
sources have circular symmetry, but the
method can be extended to deal with
asymmetric structures. Future work
will take this point into account.
We have applied these
techniques to the problem posed by 
the presence of Sunyaev-Zel'dovich effect
in multifrequency CMB maps. The best results are obtained with
matched multifilters. 
The detection level of the method is comparable
to other single-component separation methods,
but the number of assumptions and previous data 
processing is reduced to a minimum. 
Owing to the weakness of
the signals that we aim to detect, 
the determination of the amplitude of the 
clusters is biased; this is the strongest
limitation of the method. Therefore, the method
should be complemented with a posteriori
analysis (for example data fitting)
in order to improve the amplitude estimation of
weak clusters. Finally, the filters here
presented can be used in other fields of
Astronomy, such as X-ray observations,
and large-scale structure detection.

\section*{Acknowledgments}

We thank Patricio Vielva for useful suggestions and comments.
DH acknowledges support from a Spanish MEC FPU fellowship
and the hospitality of the Astrophysics Group of the Cavendish 
Laboratory during a research stay in July 2001.
RBB acknowledges financial support from the PPARC in the form of a
research grant.
RBB thanks the Instituto de F\'\i sica de
Cantabria for its hospitality during a stay in 2001.
JMD acknowledges support from a Marie Curie Fellowship of the
European Community programme {\it Improving
the Human Research Potential and Socio-Economic knowledge} 
under contract number HPMF-CT-2000-00967.
We thank FEDER Project 1FD97-1769-c04-01, Spanish DGESIC Project 
PB98-0531-c02-01 and INTAS Project INTAS-OPEN-97-1192 for 
partial financial support.

\end{document}